\documentclass{jpsj3}
\usepackage{txfonts}

\title{
Circularly polarized topological edge states derived from optical Weyl points \\
in semiconductor-based chiral woodpile photonic crystals
}

\author{
S. Takahashi$^{1,2}$\thanks{shuntaka@kit.ac.jp}, S. Oono$^3$, S. Iwamoto$^{1,4}$, Y. Hatsugai$^3$, and Y. Arakawa$^{1,4}$
}

\inst{
$^1$Institute for Nano Quantum Information Electronics, University of Tokyo, \\
4-6-1 Komaba, Maguro-ku, Tokyo 153-8505, Japan \\
$^2$Kyoto Institute of Technology, Matsugasaki, Sakyo-ku, Kyoto 606-8585, Japan \\
$^3$Graduate School of Pure and Applied Science, University of Tsukuba, \\
1-1-1 Tennodai, Tsukuba, Ibaraki 305-8577, Japan \\
$^4$Institute of Industrial Science, University of Tokyo, 4-6-1 Komaba, Maguro-ku, Tokyo 153-8505, Japan
} %\\

\abst{
The polarizations of topological edge modes in the vicinity of optical Weyl points were numerically studied in chiral photonic crystals.
We investigated two kinds of rotationally stacked woodpile structures in which planar rod arrays were vertically stacked one-by-one with an in-plane rotation angle of 60$^{\circ}$ or 45$^{\circ}$.
Both structures showed pairs of optical Weyl points having topological numbers of opposite signs for photonic bands in low orders.
Topological edge states derived from the Weyl points appeared below the light line, and were strongly confined at the air interfaces in a length shorter than the wavelength.
Their polarizations in a direction perpendicular to the propagation direction were found to be one particular circular polarization that depended on the handedness of the structural chirality.
Since these chiral photonic crystals can be fabricated using semiconductor materials such as GaAs or Si, the obtained robust planar waveguides for circularly polarized light at the interface between air and the semiconductor structure can be useful not only in photonics but also in spintronics or quantum information technology through spin-photon interfaces.
}

\begin{document}
\maketitle

%%%%%%%%%%%%%%%%%%%%%%%%%%%%%%%%%%%%%%%%%%%%%%%%%%%%%%%%%%%%%%%%%%%%%%%%%%%%%%%

A Weyl point is a single point degeneracy for two linear dispersions in three-dimensional (3D) momentum space, and is a 3D counterpart of a Dirac point in two-dimensional (2D) systems such as graphene \cite{Lu_review}.
Thus, Weyl points are expected to provide high electron mobility in 3D directions or negative magneto-resistance in electronics \cite{Burkov}, as well as large-volume single-mode lasing \cite{Abad} in photonics.
Though a Weyl point has a non-zero topological charge, like a Dirac point, the Weyl point breaks the chiral symmetry that holds for a Dirac point.
Therefore, by detuning a wavevector in any 3D direction from a Weyl point, the point degeneracy is lifted, and topological edge states are easily obtained in the energy gap.
These edge states can be applied to topologically protected waveguides without any external fields.

%%%%%%%%%%%%%%%%%%%%%%%%%%%%%%%%%%%%%%%%%%%%%%%%%%%%%%%%%%%%%%%%%%%%%%%%%%%%%%%

Since the Weyl Hamiltonian, which is a massless case of the four-spinor Dirac Hamiltonian, includes all three terms of Pauli matrices, the product of parity ($P$) and time-reversal ($T$) symmetry of the system must be broken in order to obtain Weyl points.
To break $PT$ symmetry, breaking $T$ symmetry by external magnetic fields or internal effective magnetic fields such as spin-orbit interactions is relatively difficult in practical realizations, particularly in photonics.
In the case where $T$ symmetry is maintained, possible systems forming Weyl points in a 3D reciprocal space are 3D structures with properties that break $P$ symmetry, such as chirality \cite{Lu_cal,Fan,CTChan1}.
In photonics, recent progress in 3D fabrication allows us to realize Weyl points in the microwave regime by using ceramic-based gyroids \cite{Lu_exp}, metal-coated dielectrics having twisted slots \cite{CTChan2}, or non-centrosymmetric metal inclusions in a dielectric \cite{Yang}, and in the optical regime by using polymer-based helices \cite{Noh1}.
However, these reported structures are designed for the microwave regime, or are relatively large in a sub-mm scale by using photonic bands in high orders above the light line.
In addition, the materials in those reports are not suited for scaling to or integration with current electrical/optical systems.

%%%%%%%%%%%%%%%%%%%%%%%%%%%%%%%%%%%%%%%%%%%%%%%%%%%%%%%%%%%%%%%%%%%%%%%%%%%%%%%

Here, we numerically found Weyl points in two kinds of semiconductor-based chiral woodpile photonic crystals (PhCs) consisting of planar rod arrays stacked vertically with an in-plane rotation angle of 60$^{\circ}$ or 45$^{\circ}$.
These semiconductor chiral structures, having sub-micron periodicity, can be fabricated by a micro-manipulation technique \cite{Takahashi1,Takahashi2} and are expected to work with near-infrared light.
Numerical calculations using a localized basis set \cite{Oono} were performed to obtain the section Chern number defined for a 2D section of a 3D Brillouin zone.
In both kinds of PhCs, Weyl points are found at photonic bands in low orders, and the section Chern numbers at around the Weyl points change by varying the wavevector.
Due to bulk-edge correspondence \cite{Hatsugai}, topological edge modes are found even below the light line, indicating that the edge states are well-confined at the interface between the chiral PhCs and air.
These edge states are found to be strongly polarized in one of the circular polarizations depending on the handedness of the structural chirality.
Note that though a previous report \cite{CTChan1} also showed Weyl points in a similar structure using a tight-binding method, the edge states were investigated only at metal interfaces, and their polarization was not discussed.
The obtained edge states derived from the Weyl points can be applied to topologically-protected waveguides at air interfaces, which can be integrated in the electrical/optical devices due to the use of a semiconductor material.
Furthermore, since the semiconductor-based structure can contain light sources such as quantum dots \cite{Takahashi3}, topologically-protected active optical devices such as light emitting diodes and lasers are also expected.
Considering the correspondence between circularly polarized light and electron/hole spins in the solid state, the proposed structure can be useful not only in photonics but also in spintronics or quantum information technology.

%%%%%%%%%%%%%%%%%%%%%%%%%%%%%%%%%%%%%%%%%%%%%%%%%%%%%%%%%%%%%%%%%%%%%%%%%%%%%%%

\begin{figure}[t]
\includegraphics[width=1.0\linewidth]{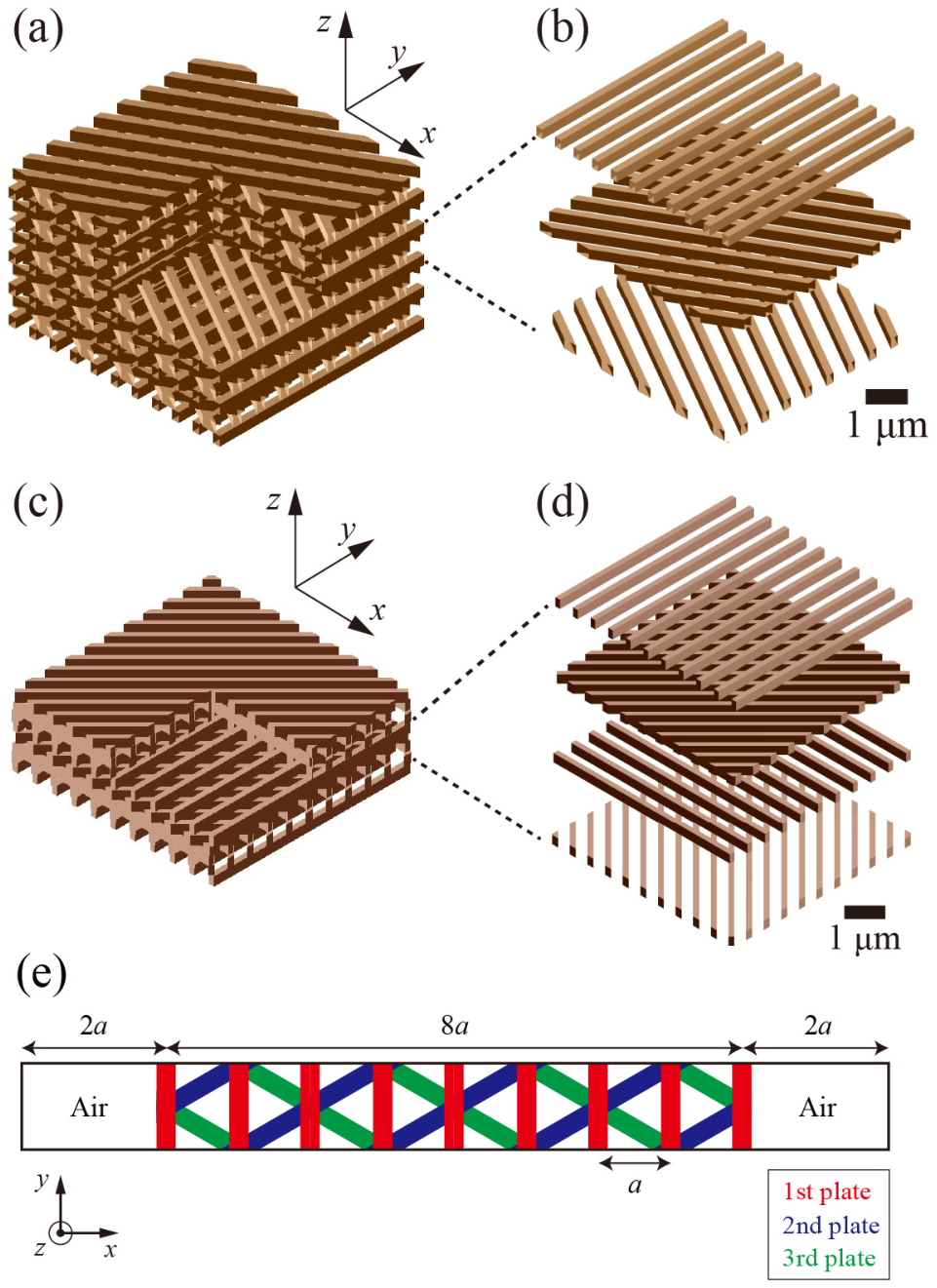}
\caption{
Schematic diagrams for the two kinds of chiral PhCs studied here.
(a) A layer-by-layer chiral PhC formed by staking plates with an in-plane rotation angle of 60$^{\circ}$.
Three plates construct a single helical unit, as shown in (b).
(c) A layer-by-layer chiral PhC formed by staking plates with an in-plane rotation angle of 45$^{\circ}$.
Four plates construct a single helical unit, as shown in (d).
The coordinates, $x$, $y$, and $z$, are defined in (a) and (c).
For clarity, a part of the structure is removed in (a) and (c), and the plates are separated from each other in (b) and (d).
(e) The calculation region for topological edge states in the PhC in (a).
Three plates constructing the helical unit are represented in different colors.
For the PhC in (c), the calculation region is consistent with (e) except for the length in the $y$-direction due to the different size of the unit cell.
}
\label{f1}
\end{figure}

%%%%%%%%%%%%%%%%%%%%%%%%%%%%%%%%%%%%%%%%%%%%%%%%%%%%%%%%%%%%%%%%%%%%%%%%%%%%%%%

One kind of 3D chiral PhC composed of plates having a rod array is shown in Fig. 1(a) and (b).
The rod pattern is rotated by 60$^{\circ}$ in the in-plane direction for each stacked plate.
Rods having a width of 130 nm and a thickness of 500/3 nm are arranged with a period of $a$ = 500 nm.
The rod material is assumed to be GaAs having a refractive index of 3.4.
These three plates construct a single helical unit having a helical pitch of $p$ = 500 nm (= $a$).
From a top view of the PhC, the structure looks a triangular lattice.

%%%%%%%%%%%%%%%%%%%%%%%%%%%%%%%%%%%%%%%%%%%%%%%%%%%%%%%%%%%%%%%%%%%%%%%%%%%%%%%

Figure 1(c) and (d) shows the other kind of chiral PhC formed by stacking similar plates with a 45$^{\circ}$ in-plane rotation.
In this structure, the width and thickness of the GaAs rods are 130 nm and 125 nm, respectively, and four plates construct a single helical unit having a helical pitch of 500 nm.
Note that the period of the rod array is 500 nm in the first and third plates, whereas it is 500/$\sqrt2$ nm in the second and fourth plates.
From a top view of the PhC, the crossing points of the rods form a square lattice.
The $x$, $y$, and $z$ axes in both PhCs are defined as shown in Fig. 1.

%%%%%%%%%%%%%%%%%%%%%%%%%%%%%%%%%%%%%%%%%%%%%%%%%%%%%%%%%%%%%%%%%%%%%%%%%%%%%%%

We performed numerical calculations using a spatially localized Gaussian basis set \cite{Oono} to obtain photonic band structures and topological Chern numbers for these chiral PhCs having periodic boundary conditions.
Note that since the Chern number is defined only in even-dimensional systems, we calculated section Chern numbers defined for a 2D section of a 3D Brillouin zone.
In the calculations of the topological edge states, the spatial calculation region is limited, as schematically shown in Fig. 1(e).
In the $x$-direction, the size of the PhCs are limited to a length of 4 $\mu$m (= 8$a$) and are sandwiched by air regions with a length of 1 $\mu$m (= 2$a$).
Both sides of the calculation region in the $x$-direction are finally terminated by perfect conductors.
The respective lengths of the calculation region in the $y$ and $z$-directions are 1/$\sqrt3$ $\mu$m (= 2$a$/$\sqrt3$) and 0.5 $\mu$m (= $a$) for the PhC with a 60$^{\circ}$ in-plane rotation, and 0.5 $\mu$m (= $a$) and 0.5 $\mu$m (= $a$) for the PhC with a 45$^{\circ}$ in-plane rotation.
Periodic boundary conditions are imposed in the $y$ and $z$-directions.
To confirm the obtained results, we also used a plane wave expansion (PWE) method and a finite-difference time domain (FDTD) method for the band structures and the edge states, respectively.

%%%%%%%%%%%%%%%%%%%%%%%%%%%%%%%%%%%%%%%%%%%%%%%%%%%%%%%%%%%%%%%%%%%%%%%%%%%%%%%

\begin{figure}
\includegraphics[width=1.0\linewidth]{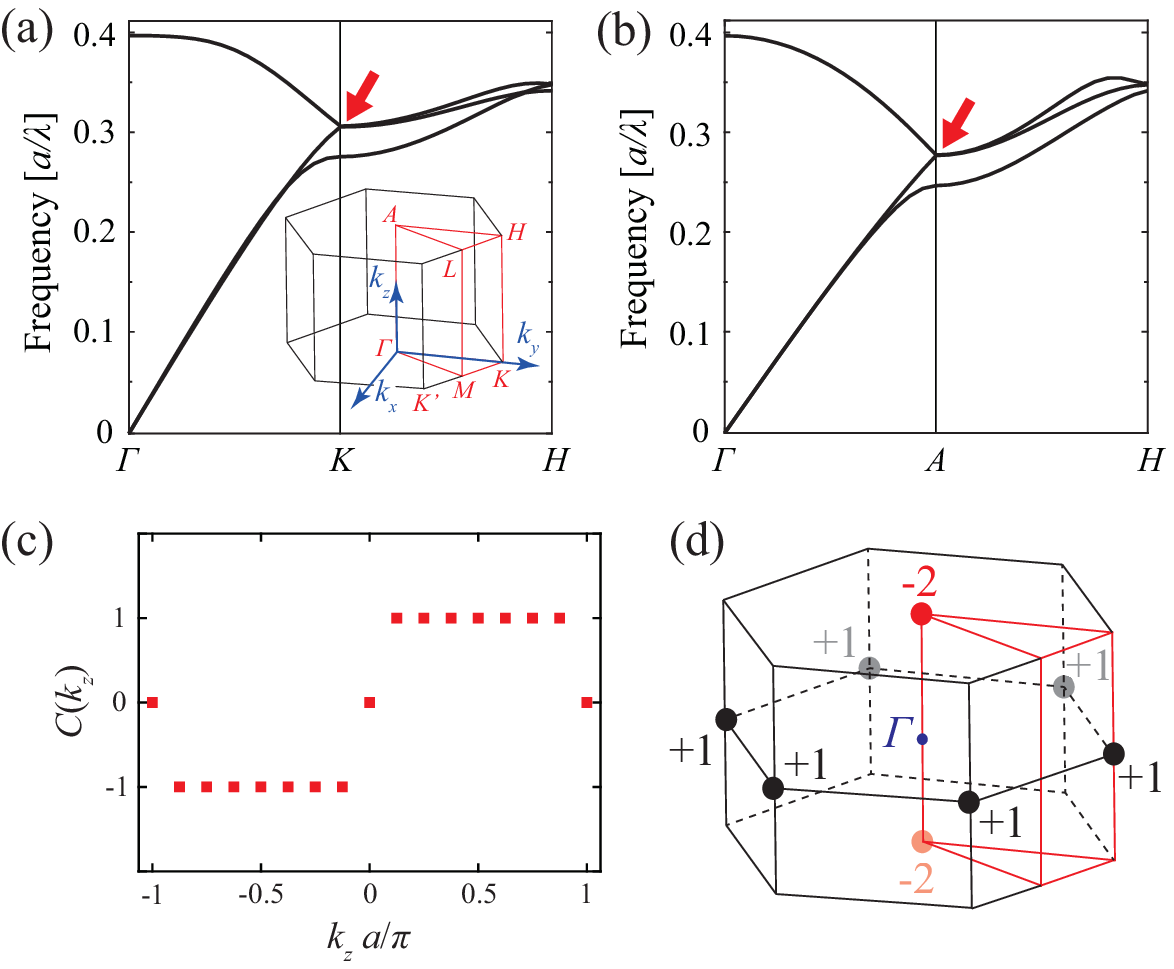}
\caption{
(a), (b) Parts of the photonic band structure for the chiral PhC with 60$^{\circ}$ in-plane rotation.
Point degeneracies in the 3D reciprocal space appear at the $K$ and $A$ points for the second and third lowest bands.
The inset in (a) shows half of the first Brillouin zone for the chiral PhC.
(c) The sum of the section Chern numbers, $C$($k_z$), for the lowest two bands.
Except for $k_z$ = 0 and $\pm$$\pi$/$a$, $C$($k_z$) shows non-zero values, indicating that the point degeneracies at the $K$ ($K'$) and $A$ points are Weyl points.
(d) Distribution of the topological charges in the first Brillouin zone.
The total topological charge integrated in the Brillouin zone is conserved to be zero.
}
\label{f1}
\end{figure}

%%%%%%%%%%%%%%%%%%%%%%%%%%%%%%%%%%%%%%%%%%%%%%%%%%%%%%%%%%%%%%%%%%%%%%%%%%%%%%%

First, we investigated the chiral PhC constructed by stacking the plates with 60$^{\circ}$ in-plane rotation.
The chirality of this PhC is right-handed.
Figures 2(a) and (b) show the photonic band structures for the lowest three bands in particular parts of the first Brillouin zone, from $\it \Gamma$ to $H$ through $K$ and from $\it \Gamma$ to $H$ through $A$, respectively.
Here, the symmetric points in the first Brillouin zone are defined in the inset in Fig. 2(a).
The wavevectors $k_x$, $k_y$ and $k_z$ point in the same directions as $x$, $y$, and $z$ in the real space.
Since the top view of the PhC looks like a triangular lattice, the second and third band show linear dispersions that are degenerate at the $K$ and $K'$ point.
This degeneracy is lifted by varying the wavevector $k_z$ along the helical axis from $K$ to $H$ in Fig. 2(a).
Calculations of the dispersions in the other reciprocal directions showed that this degeneracy occurs only at a single reciprocal point between the second and third lowest bands, indicating the existence of a Weyl point even below the light line.
The same discussion can be applied to the degenerate point at the $A$ point in Fig. 2(b).

%%%%%%%%%%%%%%%%%%%%%%%%%%%%%%%%%%%%%%%%%%%%%%%%%%%%%%%%%%%%%%%%%%%%%%%%%%%%%%%

Further evidence for these Weyl points are provided by non-zero topological numbers around the degeneracy points.
Figure 2(c) shows the sum of the section Chern numbers, $C$($k_z$), for the lowest two bands.
$C$($k_z$) has non-zero values except for $k_z$ = 0 and $\pm$$\pi$/$a$, confirming that the point degeneracies at the $K$ ($K'$) and $A$ points are Weyl points.
The value of $C$($k_z$) changes from -1 to +1 at around $k_z$ = 0, and from +1 to -1 at around $k_z$ = $\pm$$\pi$/$a$.
The sudden change by +2 at $k_z$ = 0 indicates that Weyl points having +1 charge exist at the $K$ ($K'$) points.
Regarding the change by -2 at $k_z$ = $\pm$$\pi$/$a$, on the other hand, since Weyl points must appear as a pair to conserve zero total topological charge in the first Brillouin zone, two Weyl points having -1 charge exist at the $A$ point, as shown in Fig. 2(d).

%%%%%%%%%%%%%%%%%%%%%%%%%%%%%%%%%%%%%%%%%%%%%%%%%%%%%%%%%%%%%%%%%%%%%%%%%%%%%%%

\begin{figure}
\includegraphics[width=1.0\linewidth]{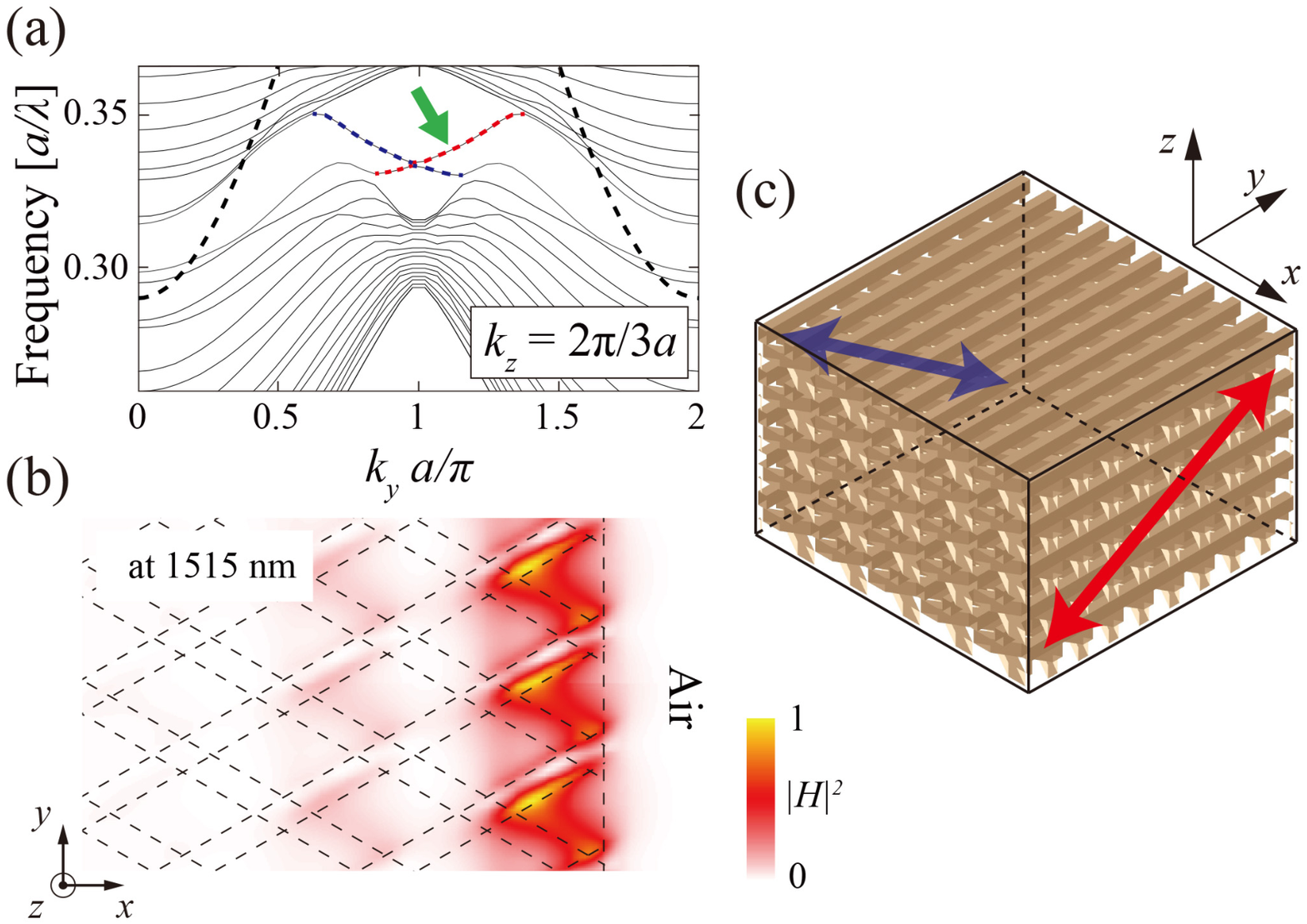}
\caption{
(a) Photonic band structure for the chiral PhC at $k_z$ = 2$\pi$/3$a$, projected in the $k_y$ direction.
Topological edge states, indicated by red and blue dotted lines for clarity, appear in the gap derived from the Weyl points.
The edge modes are below the light line, which is indicated by black dotted lines.
(b) Magnetic field intensity around the air interface for the gapless state at $a/\lambda$ = 0.33 ($\lambda$ = 1515 nm for $a$ = 500 nm) and ($k_y$, $k_z$) = (1.1$\pi$/$a$, 2$\pi$/3$a$), as indicated by the green arrow in (a).
The normalized intensity of the magnetic field is strongly localized at the interface on the right side of the PhC.
(c) Schematic diagram for about the propagation directions of the edge states.
The edge state shown in red (blue) in (a) is located in the right (left) interface, and both propagation directions are tilted in the $y$-$z$ plane.
The air interfaces are assumed only in the $x$ direction, and the structure is assumed to be infinite in the other directions.
}
\label{f1}
\end{figure}

%%%%%%%%%%%%%%%%%%%%%%%%%%%%%%%%%%%%%%%%%%%%%%%%%%%%%%%%%%%%%%%%%%%%%%%%%%%%%%%

The degeneracy at the $K$ point is resolved by detuning $k_z$ from zero, and a topologically non-trivial energy gap appears.
Due to bulk-edge correspondence \cite{Hatsugai}, topological edge states at the interface between the chiral PhC and air necessarily appear in this gap.
Figure 3(a) shows the band structure projected in the $k_y$ direction at $k_z$ = 2$\pi$/3$a$.
In this gap, clear gapless states indicated by red and blue dotted lines appear even below the light line, which is indicated by black dotted lines.
The normalized band width of these edge modes is $\Delta$$f$ / $f_c$ $\sim$ 10 \%, where $f_c$ is the central frequency of the mode dispersion.
This normalized band width is comparable to that of reported topological edge states derived from Weyl points in the microwave regime \cite{Lu_cal}.
The magnetic field intensity for the gapless state at $k_y$ = 1.1$\pi$/$a$, indicated by a green arrow in Fig. 3(a), is spatially mapped in Fig. 3(b) at the interface between the second and the third plates in Fig. 1(b).
The electromagnetic field is confined at the air interface in a length shorter than the wavelength.
For the edge state shown in blue in Fig. 3(a), the electromagnetic field is confined at the air interface on the left side of the PhC.
Therefore, the gapless states are confirmed to be topologically-protected edge states.
The propagation directions of these edge states are tilted in the $y$-$z$ plane due to the finite $k_z$, as schematically shown in Fig. 3(c).
Note that the studied structure has time-reversal symmetry as well as $C_2$ symmetry around the $x$, $y$, and $z$ axes, and the edge states propagate in both $\pm$($k_y$, $k_z$) directions.

%%%%%%%%%%%%%%%%%%%%%%%%%%%%%%%%%%%%%%%%%%%%%%%%%%%%%%%%%%%%%%%%%%%%%%%%%%%%%%%

\begin{figure}
\includegraphics[width=0.9\linewidth]{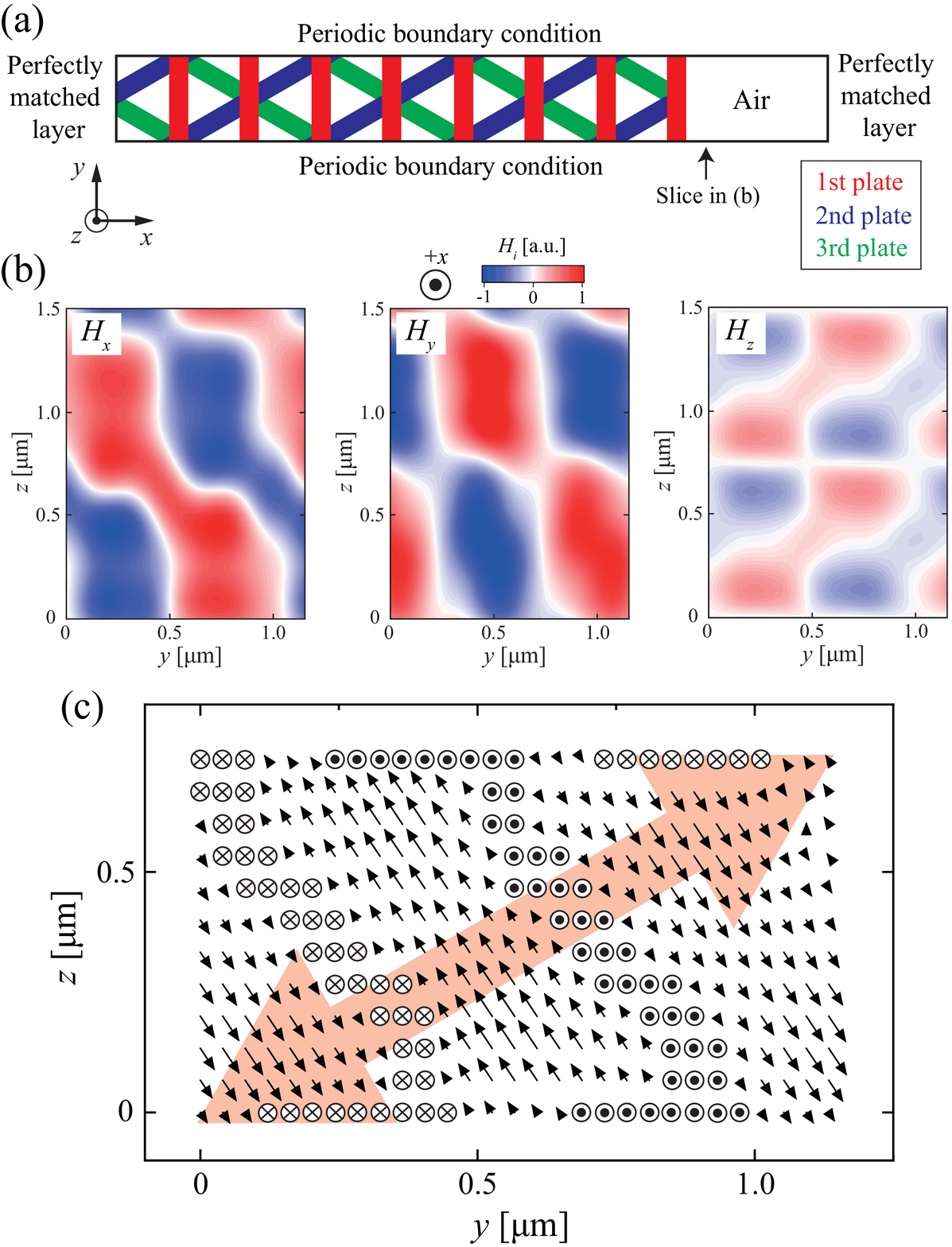}
\caption{
(a) Schematic top view of the calculated region in the FDTD method.
The region is terminated by perfectly matched layers in the $x$ direction, and by periodic boundary conditions in the $y$ and $z$ directions.
(b) Magnetic field distributions for the components of $H_x$, $H_y$, and $H_z$ at a slice in air close to the interface.
$H_x$ shifts by a quarter wavelength from $H_y$, indicating that the circular polarization component is dominant in the standing wave.
(c) Vector plot of the magnetic field component perpendicular to the propagation direction, by using the field profiles in (b).
The vectors are projected on the $y$-$z$ plane, and those having an angle more than 80$^{\circ}$ for the $y$-$z$ plane are replaced by arrows perpendicular to the $y$-$z$ plane.
The trace of the vectors forms a right-handed helix along the propagation direction indicated by a faint double-headed arrow.
}
\label{f1}
\end{figure}

%%%%%%%%%%%%%%%%%%%%%%%%%%%%%%%%%%%%%%%%%%%%%%%%%%%%%%%%%%%%%%%%%%%%%%%%%%%%%%%

The polarization of the edge modes in the steady state was then calculated using the FDTD method.
Here, we discuss the polarization component perpendicular to the propagation direction.
The PhC is assumed to be infinite in the $y$ and $z$ directions (periodic boundary conditions), and perfectly matched layers are attached in the $x$ direction, as shown in Fig. 4(a).
The perfectly matched layers are set in the PhC (air) at 4 $\mu$m = 8$a$ (1 $\mu$m = 2$a$) away from the interface.
An impulse linear dipole is set at the vicinity of the interface.
For a resonant mode at $\lambda$ = 1515 nm ($a/\lambda$ = 0.33 indicated by the green arrow in Fig. 3(a)), the magnetic field distributions in air 150 nm away from the interface are shown in Fig. 4(b).
The maximum intensity of $H_x$ is almost equal to that of $H_y$, and is two times larger than that of $H_z$.
The time evolutions of these distributions do not show any propagation, indicating a standing wave composed of two waves having ($k_y$, $k_z$) = $\pm$(1.1$\pi$/$a$, 2$\pi$/3$a$).
In the case of the two waves having linear polarization, nodes in the standing wave must be aligned at the same spatial points for all components of the field.
This is not the case in Fig. 4(b), where $H_x$ shifts by a quarter wavelength compared with $H_y$.
From these distributions, we extracted the components perpendicular to the propagation direction, ($k_y$, $k_z$) = $\pm$(1.1$\pi$/$a$, 2$\pi$/3$a$), and projected them on the $y$-$z$ plane, as shown in Fig. 4(c).
The plotted vector of the standing wave traces a right-handed helix along the propagation direction, indicating that the two waves constructing the standing wave are strongly polarized in right-handed circular polarization (RCP) \cite{Plum}.
The maximum ellipticity in Fig. 4(c) is 41$^\circ$, corresponding to a degree of circular polarization of $\sim$ 85\%.
At the air interface on the left side of the PhC, the edge state is also polarized in RCP, which is confirmed by $C_2$ symmetry around the $y$ or $z$ axis.
Note that the field component parallel to the propagation direction also shows circular rotation in the time evolution, resulting in a transverse spin \cite{Nori}.
Also, another chiral PhC having left-handed chirality also shows topological edge states propagating in the ($\pm$ $k_y$, $\mp$ $k_z$) direction at the right edge of the PhC and $\pm$($k_y$, $k_z$) at the left edge of the PhC.
This is because the sign of $C$($k_z$) is opposite for the PhCs having different chirality.
Both edge states are polarized in left-handed circular polarization (LCP).

%%%%%%%%%%%%%%%%%%%%%%%%%%%%%%%%%%%%%%%%%%%%%%%%%%%%%%%%%%%%%%%%%%%%%%%%%%%%%%%

For the other chiral PhC constructed by stacking the plates with a 45$^{\circ}$ in-plane rotation, Figs. 5(a) and (b) show the photonic band structures for the lowest six bands in the first Brillouin zone, from $\it \Gamma$ to $X$ ($Y$) and $Z$, and from $R$ to $U$ ($T$) and $S$, respectively.
Note that the chirality of the PhC is right-handed.
At the $\it \Gamma$ point, the third and fourth lowest bands are degenerated at a single reciprocal point, indicating the existence of a Weyl point.
Another point-degeneracy appears at the $R$ point, indicating another Weyl point forming a pair with the first one.
In fact, the sum of the section Chern numbers, $C$($k_z$), for the lowest three bands has a non-zero values except for $k_z$ = 0 and $\pm$$\pi$/$a$, as shown in Fig. 5(c).
The $C$($k_z$) changed from -1 to +1 at around $k_z$ = 0, and from +1 to -1 at around $k_z$ = $\pm$$\pi$/$a$.
The change by +2 (-2) at $k_z$ = 0 ($\pm$$\pi$/$a$) confirms that two Weyl points having +1 and -1 charge exist at the $\it \Gamma$ and $R$ points.
From these results, the topological charges in the first Brillouin zone are mapped in Fig. 5(d), and the total topological charge is conserved to be zero in the Brillouin zone.
Note that in Fig. 5(a), another point degeneracy appears between the fifth and sixth band at the $\it \Gamma$ point.
This degeneracy was also confirmed to be Weyl point, and another Weyl point forming a pair with the first one exists at the $S$ point in Fig. 5(b).

%%%%%%%%%%%%%%%%%%%%%%%%%%%%%%%%%%%%%%%%%%%%%%%%%%%%%%%%%%%%%%%%%%%%%%%%%%%%%%%

\begin{figure}
\includegraphics[width=1.0\linewidth]{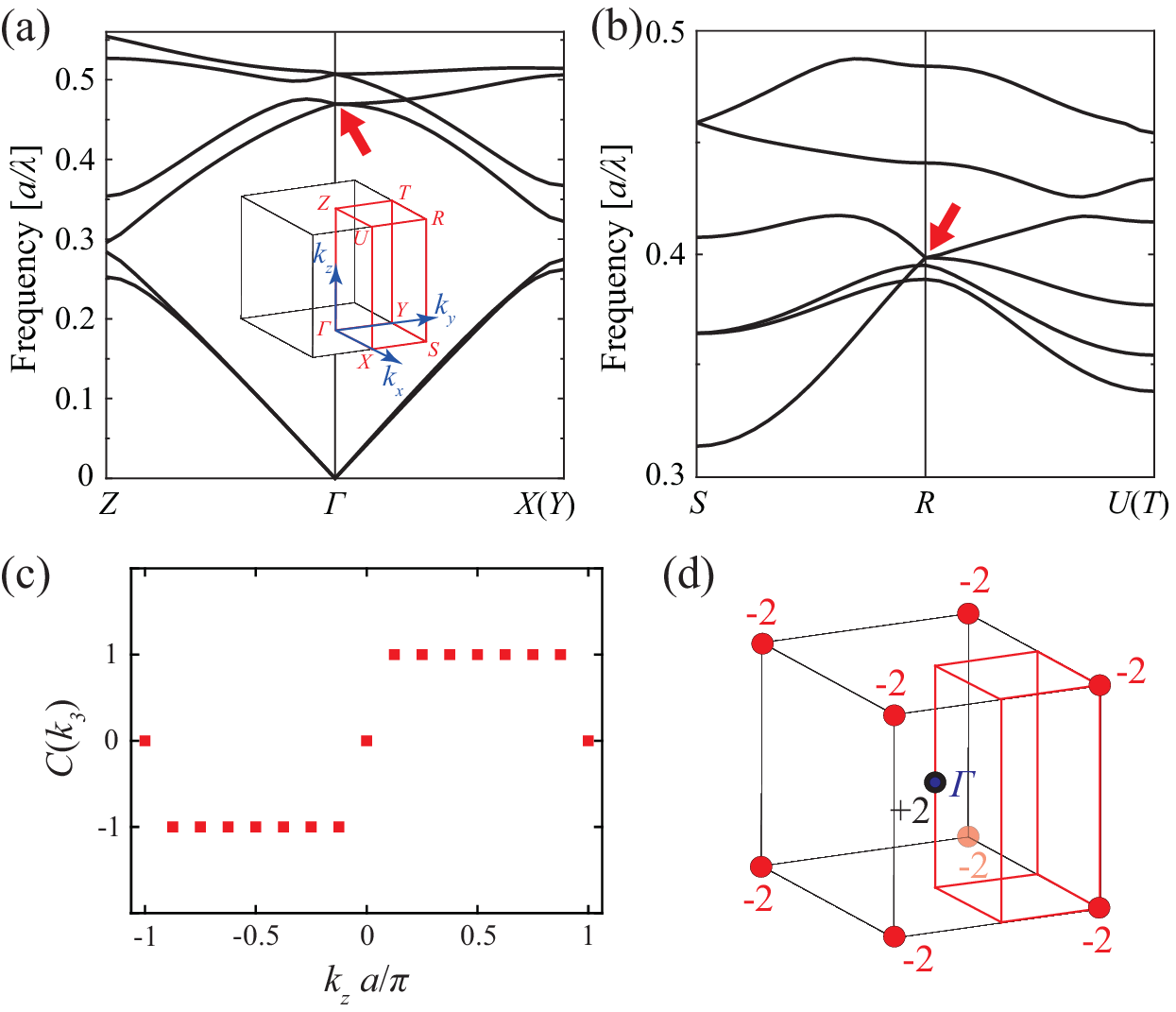}
\caption{
(a), (b) Parts of the photonic band structure for the chiral PhC with 45$^{\circ}$ rotation.
Point degeneracies in 3D reciprocal space appear at the $\it \Gamma$ and $R$ points for the third and fourth lowest bands.
Inset in (a) shows half of the first Brillouin zone for the chiral PhC.
(c) The sum of the section Chern number, $C$($k_z$), for the lowest three bands.
Except for $k_z$ = 0 and $\pm$$\pi$/$a$, $C$($k_z$) shows non-zero values, indicating that the point degeneracies at the $\it \Gamma$ and $R$ points are Weyl points.
(d) Distribution of the topological charges in the first Brillouin zone.
}
\label{f1}
\end{figure}

%%%%%%%%%%%%%%%%%%%%%%%%%%%%%%%%%%%%%%%%%%%%%%%%%%%%%%%%%%%%%%%%%%%%%%%%%%%%%%%

Topological edge states are numerically found by detuning $k_z$ from zero \cite{SM}, similar to the chiral PhC with a 60$^{\circ}$ in-plane rotation.
These gapless states are also below the light line, and the electromagnetic field is actually confined at the interface between air and the PhC.
The polarization of the edge modes is again RCP, and the sense of the circular polarization is changed in the left-handed chiral PhC.

%%%%%%%%%%%%%%%%%%%%%%%%%%%%%%%%%%%%%%%%%%%%%%%%%%%%%%%%%%%%%%%%%%%%%%%%%%%%%%%

In summary, we numerically investigated two kinds of semiconductor-based chiral PhCs consisting of planar rod arrays stacked vertically with an in-plane rotation angle of 60$^{\circ}$ or 45$^{\circ}$.
Both kinds of structures showed optical Weyl points even below the light line.
Therefore, the corresponding topological edge states appear at the air interface merely by detuning the wavevector.
By extracting the field components perpendicular to the propagation direction, we also found that the edge states are strongly polarized in one of the circular polarizations, whose handedness depended on the chirality of the PhCs.
Since the studied chiral PhCs can be fabricated using GaAs, the discovered Weyl points and the circularly polarized edge states at air interfaces are achievable in practice and will allow the realization of robust planar waveguides for circularly polarized light or topologically-protected active optical devices.

%%%%%%%%%%%%%%%%%%%%%%%%%%%%%%%%%%%%%%%%%%%%%%%%%%%%%%%%%%%%%%%%%%%%%%%%%%%%%%%

\begin{acknowledgment}

%\acknowledgment
This work was supported by Grants-in-Aid for Scientific Research, Nos. 16H06085, 26247064, 25107005, 16K13845, 17H06138, and 18K18857, Grant-in-Aid for Specially Promoted Research (15H05700), MEXT KAKENHI Grant-in-Aid for Scientific Research on Innovative Areas (15H05868), and a research grant from The Murata Science Foundation.

\end{acknowledgment}

%%%%%%%%%%%%%%%%%%%%%%%%%%%%%%%%%%%%%%%%%%%%%%%%%%%%%%%%%%%%%%%%%%%%%%%%%%%%%%%

%%%%%%%%%%%%%%%%%%%%%%%%%%%%%%%%%%%%%%%%%%%%%%%%%%%%%%%%%%%%%%%%%%%%%%%%%%%%%%%


\begin{thebibliography}{00}

\bibitem{Lu_review}
L. Lu, J. D. Joannopoulos, and M. Solja$\rm {\check{c}}$i$\rm {\acute{c}}$, Nat. Photon. {\bf 8}, 821 (2014).

\bibitem{Burkov}
A. A. Burkov, J. Phys. Condens. Matter {\bf 27}, 113201 (2015).

\bibitem{Abad}
J. B.-Abad, J. D. Joannopoulos, and M. Solja$\rm {\check{c}}$i$\rm {\acute{c}}$, Proc. Nat. Acad. Sci. {\bf 109}, 9761 (2012).

\bibitem{Lu_cal}
L. Lu, L. Fu, J. D. Joannopoulos, and M. Solja$\rm {\check{c}}$i$\rm {\acute{c}}$, Nat. Photon. {\bf 7}, 294 (2013).

\bibitem{Fan}
M. Xiao, Q. Lin, and S. Fan, Phys. Rev. Lett. {\bf 117}, 057401 (2016).

\bibitem{CTChan1}
M.-L. Chang, M. Xiao, W.-J. Chen, and C. T. Chan, Phys. Rev. B {\bf 95}, 125136 (2017).

\bibitem{Lu_exp}
L. Lu, Z. Wang, D. Ye, L. Ran, L. Fu, J. D. Joannopoulos, and M. Solja$\rm {\check{c}}$i$\rm {\acute{c}}$, Science {\bf 349}, 622 (2015).

\bibitem{CTChan2}
W.-J. Chen, M. Xiao, and C. T. Chan, Nat. Commun. {\bf 7}, 13038 (2016).

\bibitem{Yang}
B. Yang, Q. Guo, B. Tremain, R. Liu, L. E. Barr, Q. Yan, W. Gao, H. Liu, Y. Xiang, J. Chen, C. Fang, A. Hibbins, L. Lu, and S. Zhang, Science {\bf 359}, 1013 (2018).

\bibitem{Noh1}
J. Noh, S. Huang, D. Leykam, Y. D. Chong, K. P. Chen, and M. C. Rechtsman, Nature Physics {\bf 13}, 611 (2017).

\bibitem{Takahashi1}
S. Takahashi, A. Tandaechanurat, R. Igusa, Y. Ota, J. Tatebayashi, S. Iwamoto, and Y. Arakawa, Opt. Express {\bf 21}, 29905 (2013).

\bibitem{Takahashi2}
S. Takahashi, T. Tajiri, Y. Ota, J. Tatebayashi, S. Iwamoto, and Y. Arakawa, Appl. Phys. Lett. {\bf 105}, 051107 (2014).

\bibitem{Oono}
S. Oono, T. Kariyado, and Y. Hatsugai, Phys. Rev. B. {\bf 94}, 125125 (2016).

\bibitem{Hatsugai}
Y. Hatsugai, Phys. Rev. Lett. {\bf 71}, 3697 (1993).

\bibitem{Takahashi3}
S. Takahashi, Y. Ota, T. Tajiri, J. Tatebayashi, S. Iwamoto, and Y. Arakawa, Phys. Rev. B {\bf 96}, 195404 (2017).

\bibitem{Plum}
E. Plum and N. I. Zheludev, Appl. Phys. Lett. {\bf 106}, 221901 (2015).

\bibitem{Nori}
K. Y. Bliokh, F. J. Rodriguez-Fortuno, F. Nori, and A. V. Zayats, Nat. Photon. {\bf 9}, 796 (2015).

\bibitem{SM}
See Supplemental Material for characterization details and additional data.

\end{thebibliography}
\end{document}